# Predicting IoT Service Adoption towards Smart Mobility in Malaysia: SEM-Neural Hybrid Pilot Study

Waqas Ahmed[1], Sheikh Muhamad Hizam[2], Ilham Sentosa[3], Habiba Akter[4], Eiad Yafi[5], Jawad Ali[6]

UniKL Business School (UBIS), Universiti Kuala Lumpur, Kuala Lumpur, Malaysia[1, 2, 3, 4]

Malaysian Institute of Information Technology (MIIT), Universiti Kuala Lumpur, Kuala Lumpur, Malaysia[5, 6]

*Abstract*—Smart city is synchronized with digital environment and its transportation system is vitalized with RFID sensors, Internet of Things (IoT) and Artificial Intelligence. However, without user's behavioral assessment of technology, the ultimate usefulness of smart mobility cannot be achieved. This paper aims to formulate the research framework for prediction of antecedents of smart mobility by using SEM-Neural hybrid approach towards preliminary data analysis. This research undertook smart mobility service adoption in Malaysia as study perspective and applied the Technology Acceptance Model (TAM) as theoretical basis. An extended TAM model was hypothesized with five external factors (digital dexterity, IoT service quality, intrusiveness concerns, social electronic word of mouth and subjective norm). The data was collected through a pilot survey in Klang Valley, Malaysia. Then responses were analyzed for reliability, validity and accuracy of model. Finally, the causal relationship was explained by Structural Equation Modeling (SEM) and Artificial Neural Networking (ANN). The paper will share better understanding of road technology acceptance to all stakeholders to refine, revise and update their policies. The proposed framework will suggest a broader approach to investigate individual-level technology acceptance.

I. INTRODUCTION

In today's world, the term "Smart" depicts the intelligence and self-learning capabilities of non-living objects around human. The tangling of word "smart" as the smart-phone has, now, been diffused to many segments of life like smart-tv, smart-school, smart-home, smart-car, and smart-city etc. This smartness of the things is backed by numerous technologies i.e., RFID sensors, Internet of Things (IoT), Big Data Analytics and Machine Learning. The main purpose of smart objects is to provide the efficient and convenient way of living without human interference. Likewise, smart city manages the assets, resources, and services of urban areas to well-plan the dwelling, working, and commuting for inhabitants. Smart mobility is one of the main elements that enhances the smart city management by ensuring the safe, clean, and economical commuting and transportation mechanism. Smart mobility aims to cope up with numerous challenges such as traffic management, congestion mitigation, environmental impact control and infrastructure safety etc. through digital tools and techniques.

The urbanization around the world has reached to 55% while Malaysian urban density is expanded to 76% which is expected to surge at 82% in next 10 years[1]. Denizens of urban Malaysia have higher preference of purchasing car in cash on hand state [2]. Therefore Malaysian have become the 3rd highest level car ownership nation worldwide, where 93% of households own and utilize at least one vehicle [3]. The aggregate number of registered automobile is 88% (28.2 M) of total population (32.4 M) [4]. It is followed in excess use of personal vehicle over public transportation that concluded in challenges of higher congestion level, environmental impact, road safety, infrastructure impairment[2]. Due to this reason, the commuters spend more time on roads and suffer the traffic jams with higher pace as compared to the preceding years. Additionally, a survey by The Boston Consulting Group, indicated that inhabitants of Kuala Lumpur stuck in road congestion around 53 minutes daily and looking for parking spots takes their 25 minutes [5]. The situation becomes aggravate at toll plazas, particularly in vicinity of urban areas, where long queues waste the productive time and fuel, damage the road, and contaminate air quality. A smart mobility study in Klang Valley explored the alarming situation of damages due to outnumbered vehicles on roads by estimating productivity loss of RM 5 billion per annum [6]. As per World Bank report, the economic losses due to congestion across Klang Valley in 2014 were RM 20 billion, around RM 52 million daily. While stuck in traffic means being non-effective citizens, who waste the total time with the value of RM 10 billion to RM 20 billion annually by doing nothing. Similarly, the wasted fuel due to congestion surges around RM 2 billion. This excessive fuel burning poses the great environment danger and social threat. The overall price of traffic congestion in Klang Valley was projected at 1.1 to 2.2% of GDP in 2014 [7]. Being rolling down slowly on roads results in impairment of infrastructure that also costs the government. In urban areas of Malaysia, the vehicle average speed has been lowered down due to congestion that demands authorities to keep the facilities up to date [8].

The authorities have persuaded various ways to tackle such issues by penetrating the public transport network extension and ride sharing e-hailing services etc. To manage the private cars on roads the use of RFID sensor along E-Wallet has launched in 2019. This RFID tag is initially utilized for paying the e-toll or electronic toll collection (ETC) on toll plaza which is being implemented for smart parking and electronic road pricing (ERP). The RFID Tag is affixed on windscreen or headlamp of vehicle and then it is linked with E-wallet account. Previously, Intelligent Transport System (ITS) of Malaysia had implemented the infrared based On-Board-Unit (OBU) technology, SmartTag for e-tolling services in 1998. However due to lower adoption level i.e., 28% of registered vehicles, it was discontinued in August 2018. The new RFID sensor system aims to function through E-Wallet app for paying the tolls, parking fee and congestion price. It will eradicate the lag time in long queues, operates the traffic flow, manage the parking system through smart phone. Electronic road pricing (ERP) or congestion pricing will regulate the traffic by charging the vehicles during peak time. ERP will also regulate the traffic around schools, hospitals, and parks. By implementing the RFID sensors, ITS Malaysia will instigate Multilane Free Flow (MLFF) for gate-less or open-road tolling on highways. Embedding smart mobility services like RFID sensors to metropolis and enhancing its acceptance is the venture that will yield the benefits to government, citizens, and environment.

This IoT based RFID service has multiple benefits for users and government [9] but it has certain complexities that might hinder the proper implementation process. The RFID tag uses the E-Wallet payment method which requires users' personal information, bank account, debit/credit card details in order to complete the transaction. While cashless payment method in Malaysia depicted that Mobile wallet is the least used method as only 8% citizen using E-Wallet [10]. There are various studies, comprised of the technicalities of IoT enabled ETC system that work through RFID sensors, have suggested the on-going improvement and implications[11]–[14]. But there are very few studies focusing on personal and social factors of using this IoT based technology[15], [16]. Besides this, the user-based studies pertaining the motives and aspects that were backing the lack of acceptance and affected the lesser usage of SmartTag by motorists are also largely unknown. The literature on users' behavior towards transportation technology is limited in Malaysia while there is a gap in assessing mobile wallet usage for smart mobility services.

Acceptance and adoption of new technology always depend on perception and behavior of user towards the system. In era of technology advancement when Industry 4.0, Cloud computing and IoT are infused around, the concept of smart mobility is of higher priority [17]. Acceptance and continuation of innovative system such as RFID service along Mobile Wallet, is a personal choice that is mainly linked with human attitude and behavior [16]. Adopting the digital technology requires the innovative personality and digital dexterity that comprises of ability to understand the technology and passion to get benefit from it in daily activities. The main factors that play important role in users adoption are based on comfort, convenience and usefulness of technology [18]. Perceived enjoyment, trust, perceived behavior control positively impact the behavioral intention of user to utilize the ETC system[16]. As there are various service quality assessment model (SERVQUAL) in literature [19], the quality assessment of IoT based services also craves for the updated SERVQUAL model dimensions [20]. In electronic services, the matter of service quality always possesses the main factor towards customer trust, loyalty and branding of service. Personal factors like level of technological understanding and security concerns of using technology are also vigorous drivers of adopting the technology [21]. Using the Mobile Wallet service has shown the various issues of customer privacy and reluctance [22] while users privacy concerns for digital wallet are validated by extending Technology Acceptance Model (TAM) [23]. In scenario of IoT service acceptance, the intrusiveness concerns is verily undertaken in previous studies [24]. On the other hand, innovative service penetration is achieved by the tech-awareness capability of users [25]. Measuring the level of technology acceptance in various studies involved the personal understanding and usage of technology [26], [27] that describes as the digital dexterity. While social media role in disseminating the positive or negative information also intrigues the individual's behavior towards the technology use [28]. Society always backing up the human behavior in order to accept or reject the technology [29]. Society views are disseminated physically or virtually i.e., internet, both have impact on behavior to decide for technology acceptance [30]. Such antecedents of system use can better represent through research model or theoretical framework to understand the relationship towards users' adoption behavior.

The study proposes an adoption model i.e., IoT-TAM model to predict the attitude and behavior of motorist towards acceptance of Internet of Things based smart mobility service. This model aims to answer about the contextual predictors that influence the smart mobility service adoption. It will elaborate the support of existing theories in smart mobility. The model will also explore the impact of personal characteristics on behavior to adopt and use the smart mobility services from digital environment perspective. The theoretical framework will also depict the service quality model for IoT platform. The impact of circumstantial variables (i.e., digital dexterity, intrusiveness concerns, social electronic word of mouth and social norms) together with two theories in digital context (i.e., TAM and SERVQUAL) will instigate initial step towards filling the gap in determining the constructs of the smart mobility service adoption. The proposed model will be fully corroborated through a widespread data collection across Malaysian cities, which nurtures through the pilot study testified here. The outcomes will eventually support in illuminating the vital factors in adoption of smart mobility services in Malaysia.

## II. LITERATURE REVIEW

### A. Theory of Technology Acceptance Model

The research model adopted in this study is rooted in Technology Acceptance Model (TAM). TAM was proposed as a basic framework for tracing the impact of external factors on internal beliefs, attitudes and intentions of human[31]. TAM also projected that human attitude towards information systems is a consequence of two major principles as portrayed in Fig. 1. These include perceived usefulness (PU), that the concepts as an individual's work efficiency will improve by utilizing certain technology. When system facilitates work efficiency, the user apprehends the system in positive way (attitude). This optimistic attitude increases the eagerness to involve the system usability. While ease of use i.e., PEOU refers to the level of convenience in utilizing the technology. Individual perceives the PEOU as feasibility element to use the system. And these both construct effect the individual perception and form the antecedent attitude, refers as evaluative impact [32]. These both constructs build the positive or negative attitude towards intention to accept the technology. These two dimensions are nurtured by individual's exposure to external factors that are correlated with specifications of information system and the environment. TAM model directed that usage of information system is managed by behavioral intention that mutually impacted by the attitude of individual and technology usefulness [31]. However, people express the intention to accept the information system upon knowing its usefulness despite the despite the attitude towards the system. [31].

Many studies have justified Perceived usefulness as most influencing construct in TAM model as it defines the user perception towards the importance of information technology use [33]. In transportation field, perceived usefulness possesses the considerable influence on the attitude and behavior of technology users. As user acceptance behavior of electronic vehicle is supported by perceived usefulness [34]. Its prominence in effecting the behavior intention towards the adoption of vehicle navigation system is also acclaimed [35]. It has significant impact on user's attitude [36] and intention towards acceptance of e-tolling system [37]. This would allow the researchers to examine if, in smart mobility service adoption context, perceived usefulness would have any significant influence on attitude and behavioral intention. This is hypothesized as:

H1: "Perceived Usefulness (PU)" will positively effect "Attitude (ATT)" to adopt smart mobility services.

Perceived ease of use (PEOU) is described as "the degree to which a person believes that using a particular system would be free of effort" [38]. It is also the powerful determinate in shaping the user attitude by increasing the system usefulness that eventually effect the behavioral intention of the system [31]. Motorists attitude towards the usage of transport technology is backed by the perceived ease of use [39] while using the electronic tolling service attitude mostly affected by this factor [36], [37], [40]. The more convenient and easiness motorists gain from using IoT service on roads, the more satisfied they are, and therefore, the more likely they are eager to adopt the services. In perspective of this study, the author hypothesizes that perceived ease of use (PEOU) will have a positive influence on the perceived usefulness (PU) and attitude to use the IoT System. This would allow the researcher to examine if, in this context, perceived ease of use would have any significant influence. This is hypothesized as.

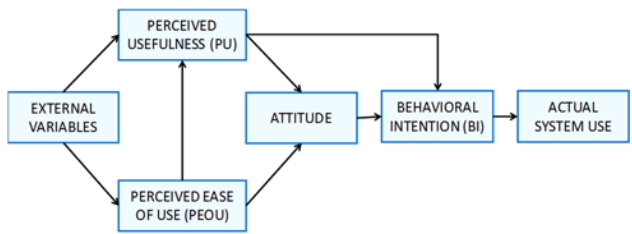

Fig. 1. Technology Acceptance Model.

H2: "PEOU" will positively effect "Attitude" to adopt smart mobility services.

H3: "PEOU" will positively effect "Perceived Usefulness" to adopt smart mobility services.

In technology adoption studies, attitude indicates positive or negative feelings of users towards system usage behavior. Theory of Planned Behavior foresees that the more fortunate an individual assesses the certain behavior, more probably he will be anticipated to perform that behavior [32]. In transportation sector it has significant relationship with behavioral intention [41] and specifically for ETC acceptance [42]. As motorists get more positive felling from using IoT service on roads, the more inclined they are, and therefore, the more likely they are eager to adopt the services. This would allow the researcher to examine if, in this context, attitude would have any significant influence on behavioral intention. This is hypothesized as:

H4: "Attitude (ATT)" will positively effect "Behavioral Intention (BI)" to adopt smart mobility services.

*B. Service Quality*

While dealing with electronic services, there is significant relationship among various users' related factors such as service quality, perceived usefulness (PU), customer's satisfaction and permeance of system usage. Service usefulness is vitally managed by service quality [43] that determines the users' behavior. In transportation research, user perception and awareness towards the service quality are of substantial importance [44]. Service quality for IoT based services can be assessed through service quality models measuring the various technological services like SERVQUAL [45], E-SERVQUAL [46] and SSQUAL [19] models.

SERVQUAL model was theorized to establish the scale of quality of services provided by organization and governments. [45] abstracted service quality in the five measurements concepts comprise (1) Reliability, (2) Responsiveness, (3) Assurance, (4) Empathy, and (5) Tangibility. This model had various extension in field of modern day technology like E-SERVQUAL, WebQual, SiteQual, IRSQ, eTailQ, PeSQ, SSTQ [19]. Later, as businesses shifted to digital services for their customer and clients through websites or web portals, the aspects of services quality updated to level of privacy of clients' data on websites, system availability, efficiency of websites and fulfilment the purpose of service. This model called as E-SERVQUAL or E-SQ [46]. It has widely used in measuring the level of service quality for websites and mobile application. An online support system for employees by government in Spain is assessed by scales of service quality (i.e., E-SERVQUAL) [47]. In Taiwan, group-purchase criteria from social media website like Facebook assessed through key quality characteristics of E-SERVQUAL [48]. Increasing labor costs have invigorated businesses to explore additional self-service alternatives that let customers to carry out services for themselves. Information technology permitted companies to practice a variety of self-service technologies (SSTs) that rise customer partaking. Self-Service Technology Quality model (SSTQUAL) was introduced by [46] that consist of 07 scales quality measures as 1) Functionality, 2) Enjoyment, 3) Security/Privacy, 4) Assurance 5) Design, 6) Convenience, 7)

Customization. However, to assess the quality of IoT based services, 04 dimension-based service quality measure is proposed that consists of privacy dimension from E_SERVQUAL and SSTQUAL, Functionality from SSTQUAL, Efficiency from SERVQUAL and Tangibility from SERVQUAL.

Several dimensions of information system (IS) service-quality have been established to asses IT success. System reliability, functionality, ease of use, system accuracy, response time turnaround time, completeness, system, flexibility, reliability, assurance and security are the various acknowledged dimension. SERVQUAL is most famous and its extensions for online or electronic business like E-SERVQUAL, SiteQual, WebQual etc. are commonly used in assessing the service quality in perspective fields [45], [46]. While currently technologies like Internet of Things, Big Data, Artificial Intelligence etc. go for the scale pertaining the measures of privacy/security, efficiency, functionality and tangibility. Service quality positively influences the constructs of TAM model [49]. Service quality model has positive impact on TAM variables in various sectors of tech-enabled services[50], [51].

In perspective of this study, the authors hypothesize that IoT-service quality will have a positive impact on the perceived usefulness, perceived ease of use, and behavioral intentions to use the IoT based RFID System. This is hypothesized as:

H5: IoT Service Quality (IoT-SQ) will positively effect "Behavioral Intention (BI)" to adopt smart mobility services.

H6: IoT Service Quality (IoT-SQ) will positively effect "Perceived Usefulness (PU)" to adopt smart mobility services.

H7: IoT Service Quality (IoT-SQ) will positively effect "Perceived Ease of Use (PEOU)" to adopt smart mobility services.

*C. Intrusiveness Concerns*

Intrusiveness concerns based on consumer's view that the service provider indecently intrudes into his or her personal life. It is assumed that when RFID technology is intricate, the privacy couldn't be absolutely guarantee, however it's interloped upon at multiple levels [52]. Perceived intrusiveness can be an obstacle to the use of custom digital services. RFID being a prevalent and ubiquitous innovation has prompted many debates because of privacy apprehensions. Towards adoption of the IoT service, intrusiveness concerns is one of the basic challenge [53]. Upon installation of app, user allows to share the a lot of information that might go for intrusiveness [54]. While intrusiveness has found significance towards the TAM variables [55]. In particular, the researchers assume that intrusiveness concerns of users will negatively impact the TAM variables i.e., perceived usefulness, perceived ease of use, and behavioral intention toward RFID tag acceptance. These hypotheses are formulated to evaluate intrusiveness as an exterior factor variable as well as to prove empirical relations with TAM model.

H8: Intrusiveness Concerns (IC) will negatively effect "Perceived Usefulness (PU)" to adopt smart mobility services.

H9: Intrusiveness Concerns (IC) will negatively effect "Perceived Ease of Use (PEOU)" to adopt smart mobility services.

H10: Intrusiveness Concerns (IC) will negatively effect "Behavioral Intention (BI)" to adopt smart mobility services.

*D. Digital Dexterity (DD)*

An individual's alertness or vigilance towards evaluating the innovation is personal innovativeness (PI). PI is the extent to which an individual perceives inclination and tendency to adopt and involve in usage of novel innovations. High level of personal innovativeness in users is projected to strengthen their positive intent towards the new innovative system or process. This can be used to classify early adopters who can either work as change mediators or else be directed specifically for adoption when resources are narrowed. To conclude, this concept of personally inclined behavior towards innovation is possibly considered to enhance broad concentrated models of IT implementation that comprised of paradigms other than individual beliefs or observations towards technology embracing decisions[26]. The term has been updated for digital environment context and this study will amend this as Digital Dexterity (DD).

The behavioral sciences with individual psychology, though, proposed that collective manipulativeness and impact and individual mannerisms for instance personal innovative behavior are possibly vital factors of technology acceptance and more imperative component in prospective users' assessment to use [56]. [57] said that individuals with higher commitment towards the innovation is backing the very success of the innovation process. The construct determines the innovative behavior of an individual in a scale from high to low, hence assisting to recognize individuals who are ascent to accept innovations before or after than others [58]. In study of an information technology associated to new innovations, it is essential to discover individual personal innovativeness towards the innovation. It is an important predictor or anticipator towards technology acceptance [59]. While in transportation studies, it also showed significance and positive relationship with TAM variables like perceived ease of use, perceived usefulness, attitude and behavioral intention to make use of the technology [60]. In accordance with the above, this study contends that drivers with higher level of innovativeness will positively inclined towards the RFID tag, therefore we hypothesis that:

H11: Digital Dexterity (DD) will positively effect "Perceived Ease of Use (PEOU)" to adopt smart mobility services.

H12: Digital Dexterity (DD) will positively effect "Perceived Usefulness (PU)" to adopt smart mobility services.

H13: Digital Dexterity (DD) will positively effect "Behavioral Intention (BI)" to adopt smart mobility services.

*E. Social Electronic Word of Mouth (Social-eWoM)*

Social electronic word of mouth is an interaction among online users by social interactive websites has proved to be the most preferred medium of electronic word of mouth (eWOM) system. [61]. Sharing and collaborating with consumers via social networking websites such as Facebook, twitter, etc. facilitate customers with probably unbiased information of products on anonymity base [62]. Electronic word of mouth has substantial effect on traveling sector. It is considered to be the valuable means of information influencing the travelers' behavior regarding their plans [63]. It is discovered that users frequently get online reviews submitted or shared by other tourists are conversant, pleasant, and trustworthy than communicated by journey facilitator organization [64]. Importance of dispersion of information through electronic mean for travelling distinguished by [65]. This eWoM proved significant impact on TAM variables in acceptance of ETC technology [40]. As drivers know more about this IoT based technology over internet they will tend to experience this. In this study perspective, it is hypothesized that social EWOM will have a positive impact on perceived usefulness, perceived ease of use, attitude and behavioral intentions to use the IoT based RFID ETC System. This is hypothesized as:

H14: "Social Electronic Word of Mouth (SWoM)" will positively effect "Intention (BI)" to adopt smart mobility services.

*F. Subjective Norm (SN)*

Higher the realization from other people's influence i.e., subjective norms, the more inclination by individual's attitude towards performance. [66]. Researchers depicted that more socially influenced individuals are likely to adopt certain technology. Subjective norms proved to be one of the key factors with extended TAM in numerous papers strengthening the user behavior towards the studied system and found positive and significant relationship with TAM variables [35]. In same way, society also firmed its impact on individual in transportation technology acceptance studies. Studies pointed out the social influence role in acceptance of E-tolling technologies in Taiwan and Indonesia [57].

The more socially influenced opinions motorists get for using IoT service on roads, the more indulged they are towards the technology, and therefore, the more likely they are eager to adopt the services. The authors hypothesize that subjective norm will have a positive impact on behavioral intention to utilize the IoT based RFID System. This would allow the researchers to examine if, in this context, subjective norm would have any significant influence. This is hypothesized as:

H15: "Social Norm (SN)" will positively effect "Intention (BI)" to adopt smart mobility services.

*G. Theoretical Framework*

A Theoretical framework is established after evaluating the well-known models in the domain of technology adoption. The model has origin in Technology Acceptance Model, and it is extended with new concepts. The proposed research framework comprises of five independent variables i.e., digital dexterity, perceived intrusiveness, service quality for Internet of Things services, social electronic word of mouth and subjective norm while there are four dependent variables i.e., perceived ease of use, perceived usefulness, attitude and behavioral intention. The research model is portrayed in Fig. 2, the shaded area representing the classical Technology acceptance model (TAM) variables.

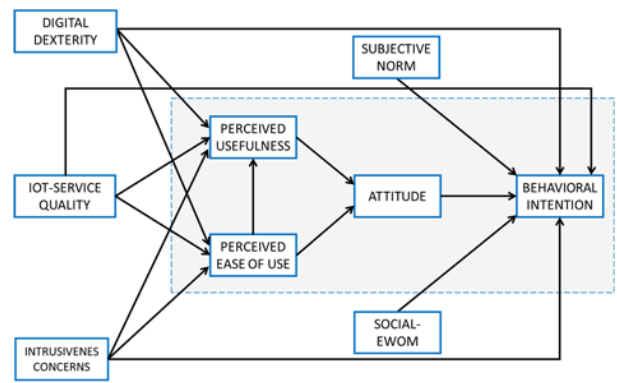

Fig. 2. Proposed Theoretical Framework (IoT-TAM).

III. METHODOLOGY

This study is based on assessment of digital factors' impact on human behavior therefore philosophy of the research is positivism. In positivism, the causal research is undertaken which deals with the cause and effect analysis by using the quantitative data. The research approach of this quantitative paper is deductive that deals with the adoption of established theory and then extends it as per the study requirement. TAM and SERVQUAL theories are used and extended in smart mobility context. Research strategy of study is Survey by which the data has collected through mono-method i.e., quantitative data technique with cross-sectional approach of time horizon [67].

Primary data for this pilot study was collected through survey. A research questionnaire was adopted from previous studies and developed for the smart mobility context. Questionnaire instruments for technology acceptance model (TAM) were taken from [38], [68] and service quality items were adopted from [19], [45], [46]. Other factors instruments were adopted as digital dexterity [26], [69], intrusiveness concerns [55], social electronic word of mouth [70] and subjective norm [71]. The questionnaire was divided into 03 section. First section focused on demographic variables, section 02 pertained the variables questions and third section consisted of respondents view on RFID system utility. The section 02 was based on measures smart mobility adoption factors such as digital dexterity, Social EWOM, Service quality, perceived usefulness, perceived ease of use, attitude, and subjective norm. To assess the response towards acceptance of smart mobility service, a five-point Likert-scale was utilized. The scale scored as range from 1 = "strongly disagree" to 5 = "strongly agree".

The analysis for pilot study mainly focuses the validity and reliability of questionnaire instruments which paves the way for full study data analysis. However, this paper has undertaken the SEM-Neural analysis approach for pilot data. To analyze the reliability of questionnaire items, Cronbach alpha, was measured through SPSS 25. Composite reliability, factor loading, discriminant validity, HTMT and r-squared values were analyzed through SmartPLS tool. Structural modeling of proposed framework was also measured by SEM in SmartPLS. Furthermore, Normalized importance of predictors towards the dependent variable (intention to adopt RFID services) were measured by Neural Network in SPSS25.

## IV. RESULTS

### A. Descriptive Results

To validate the research model, a pilot survey was performed in Klang valley. The terms Klang Valley is used for the geographical location based on vicinity of Kuala Lumpur district, Selangor District, Federal territories, and surrounding areas. It is also known as Greater Kuala Lumpur and its population is 7.8 million which is 25% of country population [72]. A field survey was organized in this region to collect the response from drivers. As initially RFID service is used for paying toll on highways therefore rest areas on highways were considered for data collection point as illustrated in Fig. 3. Three points on highways were taken for the survey as depicted in Table I.

The survey process was completed in 04 weeks and responses were collected through face to face survey. The questionnaires were printed in both English and Malay language. The respondents were introduced with the purpose of study prior to getting their views. Total 80 responses were collected for this pilot study which are suitable for required analysis. The description of respondents according to section 1 of questionnaire i.e., demographic variables are described in Table II. The particulars of demographic data showed the more participation of male, private sector employees, age group 31 to 40, monthly income RM5000 to RM8000, single car ownership and daily toll use frequency.

TABLE. I. SURVEY POINTS

| Point | Location Name |
|---|---|
| 1 | OBR - Sungai Buloh (PLUS Expressways) |
| 2 | R&R - Awan Besar (KESAS Highway) |
| 3 | OBR - USJ21, Subang Jaya (ELITE Highway) |

TABLE. II. DEMOGRAPHIC RESULTS

| Gender | % | Occupation | % |
|---|---|---|---|
| Male<br>Female | 84<br>16 | Govt Sector<br>Private Sector<br>Student<br>Business<br>Others | 10<br>51<br>14<br>20<br>5 |
| **Age** | **%** | **Income Level** | **%** |
| 18-30<br>31-40<br>41-50<br>>51 | 28<br>42<br>21<br>9 | Below RM 2,000<br>RM 2,000 – RM 5,000<br>RM 5,001 – RM 8,000<br>RM 8,001 – RM 15,000<br>More Than RM 15,000 | 4<br>31<br>48<br>16<br>1 |
| **Race** | **%** | **Education** | **%** |
| Malay<br>Chinese<br>Indian | 74<br>15<br>11 | Diploma/Certificate<br>Primary School<br>Secondary School<br>University<br>Others | 6<br>6<br>33<br>54<br>1 |
| **Vehicle Ownership** | **%** | **Toll Use Frequency** | **%** |
| 0<br>1<br>2<br>3 | 3<br>62<br>29<br>6 | Daily<br>Weekly<br>Monthly | 76<br>20<br>4 |

### B. Preliminary PLS-SEM

To assess the instrument validity and reliability of measuring in such way that the questionnaire items representing the variables are measuring the characteristics of variables. The tests of reliability to assess internal consistency (i.e., Cronbach alpha reliability and composite reliability) and validity (construct validity, discriminant validity) are performed. For this purpose, numerous analysis i.e., Cronbach Alpha Reliability test, Composite Reliability, Average Variance Extracted (AVE), Discriminant Validity HTMT Test were performed on SmartPLS. The recommended value of Cronbach Alpha is classified as excellent when higher than 0.9, strong when higher than 0.8 and acceptable when higher than 0.7. In case of its value less than o.7, the instruments have to recheck for exclusion of non-reliable items. Similarly, composite reliability is another internal measure of items consistency and its threshold value is 0.7 which is achieved by this pilot study analysis. These reliability tests will ensure the model is fit for causal analysis. In Table III, all the constructs are fulfilling the recommended criteria of reliability test as Cronbach alpha value of all variables is higher than 0.8 and some constructs have more than 0.9. It means the model variable are internally consistent for further analysis [73]. In next step, convergent validity was measured by average variance extracted (AVE) and its minimum accepted value was 0.5. The value AVE in Table III, shows all constructs are valid. However, Service quality has the value less than 0.5 because it is second order construct. While validity is the scale used for only first order construct measurement. Validity tests ensures there are no corelated variables in the model who are demonstrating the same characteristics. Discriminant validity is one the most important test for guaranteeing the success of pilot study. There are numerous methods used for discriminant validity like Fornell-Lacker Criterios, HTMT etc. This paper has undertaken the HTMT (Heterotrait-Monotrait Ratio of Correlations Test). The general rule of correlation among model factors that the value between two variables should be less than 0.9, the same rule is applied in HTMT measure. In Table IV, the discriminant validity values are accomplishing the set standards. There are no highly co-related variables in the construct set that needs to exclude from the model.

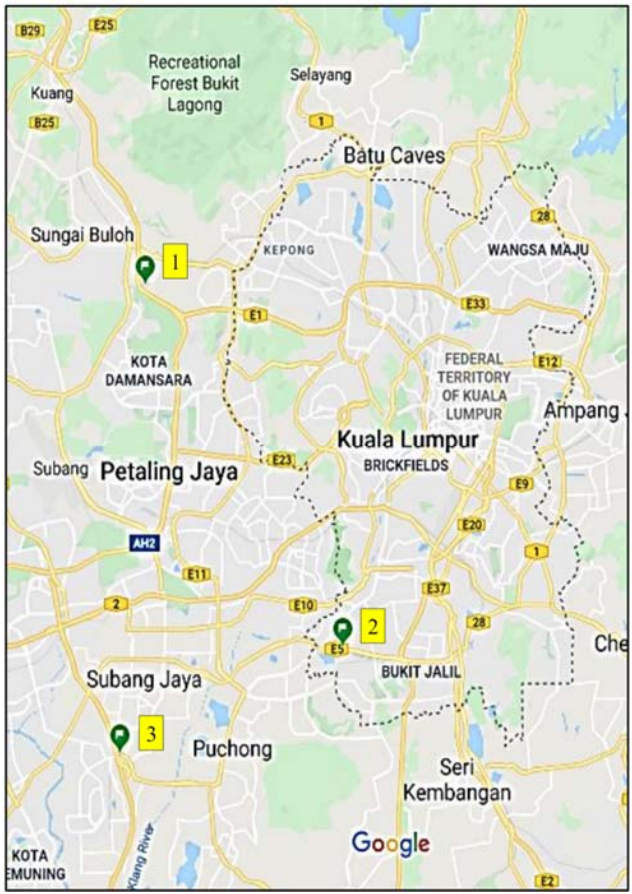

Fig. 3. Survey Location Point (Klang Valley).

The proceeding analysis of pilot survey was based on outer loading. In this phenomenon, the contribution of items or questions towards their respective construct is measured. The standard acceptance value for outer loading is more than 0.7. In Table V, indicators or items of respective constructs are shown by their respective outer loadings. Here some items had not met the criteria i.e., value should be greater than 0.70 [74]. Therefore, items from Perceived ease of use (PE5), Digital dexterity (DD4), Social electronic word mouth (WM3, WM4, WM5) were deleted from model due to the outer loading value less than 0.70. In Table V, outer loading values are normal and showing the accepted impact level.

*C. SEM-Neural Analysis*

Towards causal structuring of model, preliminary analysis was conducted internally between the construct and respective items. The next step of analysis was to assess the causal relationship between the model constructs therefore Structural Equation Modeling (SEM) and Artificial Neural Networking (ANN) were employed. However, for pilot study with small sample size the SEM and ANN are not highly validated in terms of assessing detailed impact of relationships, data normality, RMSE etc. This pilot survey aims to pave the way for full scale study. But these tests can well-predict the behavior through overall variance in SEM and normalized importance in ANN while neglecting the sample size[75], [76].

TABLE. III. RELIABILITY AND VALIDITY

| Variables | Items | Cronbach Alpha | rho_A | Composite Reliability | AVE |
|---|---|---|---|---|---|
| Attitude (ATT) | 04 | 0.938 | 0.939 | 0.956 | 0.844 |
| Behavioral Intention (BI) | 05 | 0.964 | 0.964 | 0.972 | 0.874 |
| Digital Dexterity (DD) | 06 | 0.894 | 1.022 | 0.907 | 0.625 |
| Efficiency (EF) | 04 | 0.897 | 0.898 | 0.928 | 0.763 |
| Functionality (FN) | 04 | 0.822 | 0.828 | 0.882 | 0.653 |
| Intrusiveness Concerns (IC) | 04 | 0.928 | 0.932 | 0.949 | 0.824 |
| IoT-Service Quality (SQ) | 17 | 0.924 | 0.929 | 0.934 | 0.451 |
| Perceived Ease of Use (PEOU) | 04 | 0.933 | 0.934 | 0.952 | 0.832 |
| Perceived Usefulness (PU) | 05 | 0.957 | 0.959 | 0.967 | 0.855 |
| Privacy (PR) | 05 | 0.897 | 0.898 | 0.924 | 0.711 |
| Social EWOM (SWOM) | 03 | 0.865 | 1.152 | 0.879 | 0.551 |
| Subjective Norm (SN) | 04 | 0.911 | 0.912 | 0.938 | 0.790 |
| Tangibility (TG) | 04 | 0.809 | 0.810 | 0.876 | 0.639 |

TABLE. IV. DISCRIMINANT VALIDITY - HTMT

|  | ATT | BI | EF | FN | IC | PEO | PU | DD | PR | SWOM | SN | TG |
|---|---|---|---|---|---|---|---|---|---|---|---|---|
| ATT |  |  |  |  |  |  |  |  |  |  |  |  |
| BI | 0.782 |  |  |  |  |  |  |  |  |  |  |  |
| EF | 0.502 | 0.526 |  |  |  |  |  |  |  |  |  |  |
| FN | 0.304 | 0.399 | 0.76 |  |  |  |  |  |  |  |  |  |
| IC | 0.511 | 0.667 | 0.41 | 0.29 |  |  |  |  |  |  |  |  |
| PEOU | 0.83 | 0.791 | 0.551 | 0.403 | 0.614 |  |  |  |  |  |  |  |
| PU | 0.812 | 0.785 | 0.591 | 0.381 | 0.52 | 0.862 |  |  |  |  |  |  |
| DD | 0.239 | 0.423 | 0.417 | 0.374 | 0.43 | 0.25 | 0.26 |  |  |  |  |  |
| PR | 0.247 | 0.421 | 0.65 | 0.643 | 0.249 | 0.342 | 0.35 | 0.502 |  |  |  |  |
| SWOM | 0.307 | 0.395 | 0.447 | 0.408 | 0.303 | 0.361 | 0.302 | 0.656 | 0.524 |  |  |  |
| SN | 0.654 | 0.767 | 0.318 | 0.176 | 0.419 | 0.638 | 0.646 | 0.264 | 0.265 | 0.416 |  |  |
| TG | 0.098 | 0.18 | 0.564 | 0.702 | 0.075 | 0.199 | 0.165 | 0.415 | 0.461 | 0.29 | 0.145 |  |

TABLE. V. OUTER LOADING

| | Items | Outer Loading | | Items | Outer Loading | | Items | Outer Loading |
|---|---|---|---|---|---|---|---|---|
| Attitude | AT1 | 0.893 | Intrusiveness Concerns | IN1 | 0.879 | Privacy | PR5 | 0.762 |
| | AT2 | 0.92 | | IN2 | 0.929 | Perceived Usefulness | PU1 | 0.893 |
| | AT3 | 0.926 | | IN3 | 0.931 | | PU2 | 0.922 |
| | AT4 | 0.935 | | IN4 | 0.891 | | PU3 | 0.939 |
| Behavioral Intention | BI1 | 0.889 | Perceived Ease of Use | PE1 | 0.905 | | PU4 | 0.957 |
| | BI2 | 0.929 | | PE2 | 0.882 | | PU5 | 0.91 |
| | BI3 | 0.947 | | PE3 | 0.91 | Subjective Norms | SN1 | 0.848 |
| | BI4 | 0.943 | | PE4 | 0.908 | | SN2 | 0.922 |
| | BI5 | 0.965 | Digital Dexterity | DD1 | 0.742 | | SN3 | 0.891 |
| Efficiency | EF1 | 0.874 | | DD2 | 0.782 | | SN4 | 0.894 |
| | EF2 | 0.882 | | DD3 | 0.861 | Tangibility | TG1 | 0.717 |
| | EF3 | 0.863 | | DD5 | 0.892 | | TG2 | 0.78 |
| | EF4 | 0.876 | | DD6 | 0.862 | | TG3 | 0.852 |
| Functionality | FN1 | 0.854 | Privacy | PR1 | 0.888 | | TG4 | 0.841 |
| | FN2 | 0.818 | | PR2 | 0.841 | Social Electronic Word of Mouth | WM1 | 0.785 |
| | FN3 | 0.757 | | PR3 | 0.87 | | WM2 | 0.806 |
| | FN4 | 0.799 | | PR4 | 0.848 | | WM6 | 0.874 |

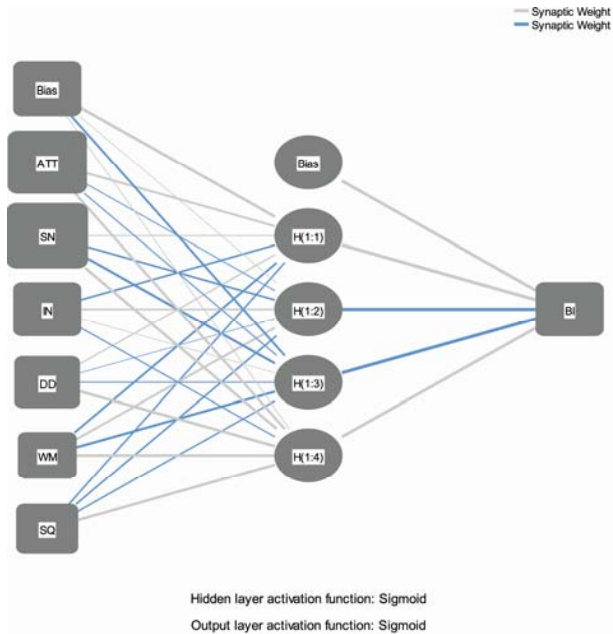

Fig. 4. ANN Diagram.

In SPSS 25, we conducted the Neural Networks test by multilayer perception tool. The exogenous variables, who had a direct impact by hypothesis on dependent variable (Behavioral Intention) were implied in covariate section with standardized rescaling. Partition of dataset was based on 70% training and 30% testing. The architecture was customized with one hidden layer. Sigmoid function was applied in both hidden layer and output layer with normalization correction 0.02. Batch training with optimization algorithm was applied as scaled conjugate gradient. Fig. 4 is depicting the synaptic weight of independent variables towards dependent variable. Normalized importance is the main tool for our study, as it shows the most important factor in terms of predicting the outcome of dependent variable. In our case, Subjective Norm (SN) is the best predictor followed by Attitude (ATT) with the highest rank. The least power of predictor is social electronic word of mouth (WM) while intrusiveness concern (IN), digital dexterity (DD) and service quality (SQ) have moderate prediction approach as presented in Fig. 5. The final output from neural network is prediction of scoring the behavioral intention on 5-point likert-scale towards adoption of smart mobility service. In Fig. 6, the chart showed the maximum points from agree to strongly agree scale.

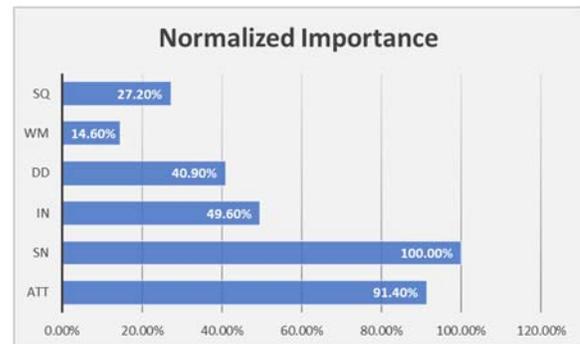

Fig. 5. redictors Normalized Importance.

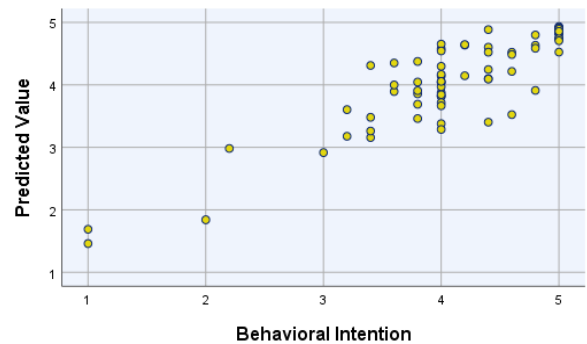

Fig. 6. Dependent Variable Prediction.

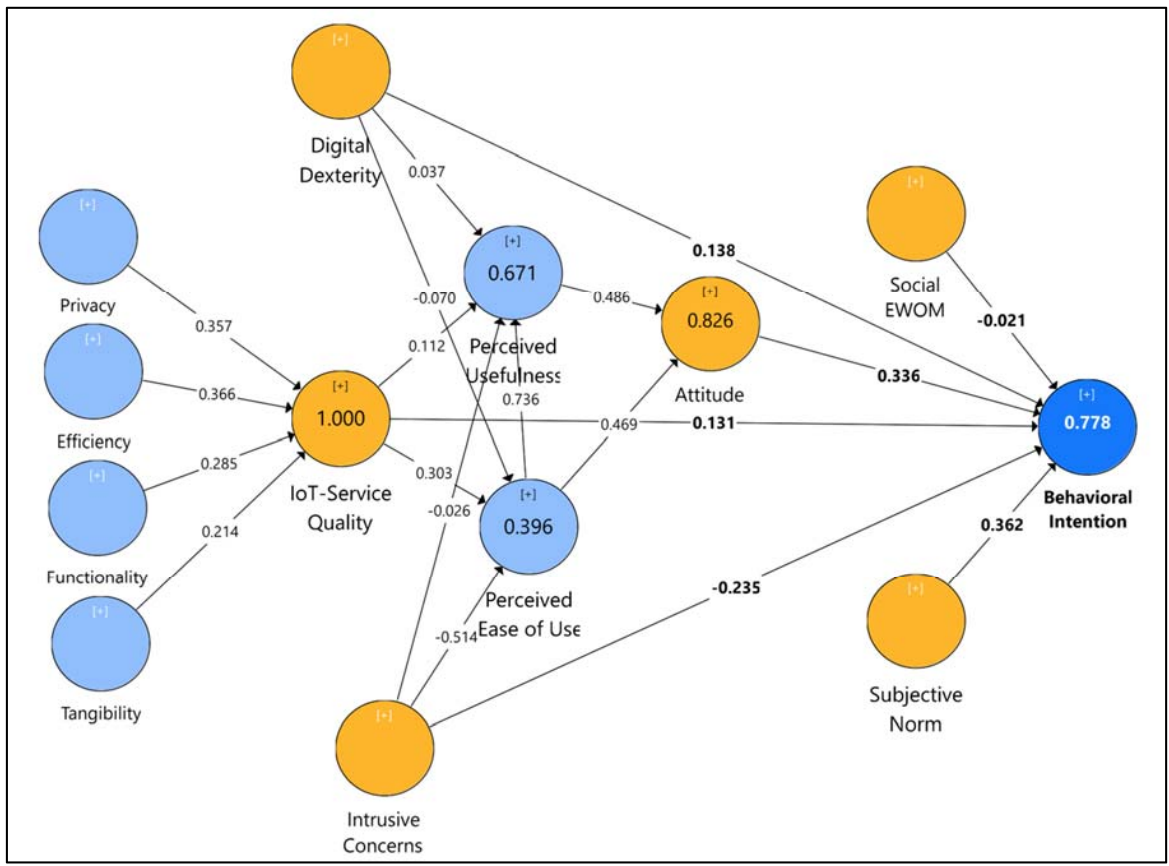

Fig. 7. Structural Equation Modeling (SEM) Causal Model.

SEM was also applied through SmartPLS software to realize the importance of constructs towards adoption of RFID service. SEM has multiple type of analysis i.e., Variance and Covariance based analysis. Although, variance-based analysis method used for pilot study data. The structural equation modeling of proposed model is illustrated in Fig. 7. The overall variance value is 0.778 that means factors in this model can predict 77.8% impact on behavioral intention towards the adoption of RFID service. The regression weights of all direct variables to BI are significant except SWOM. Subjective Norm emerged with the highest impact on BI (0.364). Attitude is appeared as significant influencer with 0.336 weight. Intrusiveness concerns showed the strong but negative impact on behavioral intention as hypothesized earlier. However, the least and negligible impact appeared from Social EWOM. The SEM analysis shows the model is well constructed with support of proposed hypothesis. As our all proposed hypothesis statements are proved true except one (i.e., Social EWOM and BI) in the preliminary results. This hypothesis will be considered to not include in final study or will identify the items to correct it [77].

## V. CONCLUSION

The aforementioned analytical findings demonstrate the validation of our proposed research model. The results of instruments' reliability, validity, correlation, and outer loading evaluation proved the pilot survey authenticated for final study. The SEM-Neural approach has described the predictors of model that will support in finalizing the model for widespread survey. The relative importance in neural network is considered as major tool to take or drop the predictor in model. Merely relying on it does not guarantee that model is accurate or not. However, using SEM with Neural can explain the accuracy of model. As the least important predictor in Neural (i.e., WM) is appeared as insignificant in SEM. To ensure the precision and accuracy of the framework, SEM -Neural approach is recommended. In SEM approach, the impact of predictors on TAM variables (PU & PEOU) is considerable. The impact of main constructs (PU and PEOU) of TAM model on Attitude has shown the high variance of 0.826. The PU was hypothesized by 03 predictors (DD, IN, SQ) and variance calculated due to impact of these was 0.671 that is considered as higher. The empirical conclusion has resulted in reliable, valid and accurate theoretical framework for smart mobility service adoption in Malaysia.

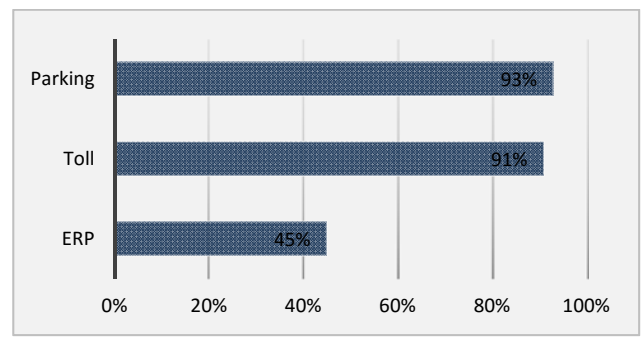

Fig. 8. RFID user Preference.

There is a huge potential for the growth of IoT based services in Malaysia, although such services are currently at initial stage. This pilot survey has concluded that higher number of drivers i.e., 93% are willing to use RFID for parking purpose. Use of RFID for tolling is preferred by 91% of respondents. While inclination to use RFID for Electronic Road Parking (ERP), only 45% respondent showed willingness as illustrated in Fig. 8. At this initial stage, the users' behavior to utilize RFID is portraying the image of future adoption pattern. The study will support authorities to devise the future policies regarding penetration of the RFID system. It will help in transforming the traditional transport system to Intelligent Transport System (ITS) with help of connected vehicles of RFID tag. It will also support in controlling congestion issue on urban roads. Furthermore, the Mobile-Wallet cashless payment method will integrate to make the things smart and intelligent. The given facts and figures of behavior preference towards smart mobility services will support the policymakers in service structure formation. It will also contribute to academic literature on smart mobility and Internet of Things in human computer interaction. It will assist the industry specific personnel for better understandings of user preferences. Lastly, the paper will contribute to the enduring research into finest technology adoption models that are pertinent for advancing the smart city contexts.

In this paper, TAM was extended with IoT and Digital environment perspective. The constructs IoT-Service Quality was introduced in this paper which had no traces in prior studies. To well-understand the human behavior in digital world, this paper introduced the new construct of Digital Dexterity in human-computer interaction literature. Practically, all pertained literature is cohered with the aspects that affect a users' decision to embrace IoT services with higher tendency of TAM. Furthermore, territory specific studies particularly in Malaysia, about the reasons that take the lead to the acceptance of Internet of Things services are almost scarce. A conceptual study was essential to comprehend the fundamental considerations and validate the analysis tools for full scale survey that encourage users to utilize IoT service. This paper emphasized the significance of various aspects on consumer intention to use IoT and to suggest it by propositioning a model by bringing into account technology adoption model and by expanding it with numerous new constructs. The proposed theoretical model is a provision to the existing literature since several uncharted variables have been included that will affect behavioral intent to make use of IoT services. This paper features the significance of pilot study regarding contribution to Information System literature and smart mobility context.

We propose to apply the validated instrument from this pilot study, conducted in Klang Valley, to the widespread survey with the large sample size across the main highways of Peninsular Malaysia. The results from the full-scale study will potentially contribute to the development of a context-specific model that can be employed to investigate smart mobility adoption and other technology diffusion in emerging economy. There is still unexplained variance in the model, that can be explained by adding new variables in the proposed framework. The framework has not encapsulated the system features in determining adoption behavior that could significantly impact the behavioral intention. As the research investigates RFID usage, the further studies could include the nature of RFID i.e., Passive RFID and Active RFID system in usability assessment. The hybrid analysis of SEM and Neural Network is the pioneer method in preliminary data analysis. However various empirical limitations of pilot survey such as covariance-based SEM analysis, detailed Neural Network processes and data normality factors can be overcome by large sample size. In future studies, various hybrid analysis techniques for predicting behavior in a pilot study such as System Dynamic Analysis (SDA), Association rule, fuzzification etc. could predict the more dynamic inferences.

REFERENCES

[1] United Nations, "World Urbanization Prospects 2018 Highlights," 2018.

[2] A. S. Abdelfatah, M. Z. Shah, and O. C. Puan, "Evaluating the Sustainability of Traffic Growth in Malaysia," J. Traffic Logist. Eng., vol. 3, no. 1, pp. 6–11, 2015.

[3] Asia IoT Business Platform, "Mobility: A key aspect in Malaysia's Smart City Plans - Asia IoT Business Platform," 2019. [Online]. Available: https://iotbusiness-platform.com/blog/mobility-a-key-aspect-in-malaysias-smart-city-plans/. [Accessed: 30-Dec-2019].

[4] MAA, "Malaysian Automotive Association," 2017. [Online]. Available: http://www.maa.org.my. [Accessed: 25-Feb-2018].

[5] BCG, "Unlocking Cities: The impact of ridesharing in Southeast Asia and beyond 2017," 2017.

[6] D. Hong and L. W. Wong, "A study on smart mobility in Kuala Lumpur," Proc. 2017 2nd Int. Conf. Comput. Commun. Technol. ICCCT 2017, pp. 27–32, 2017.

[7] World Bank Group, "Malaysia Economic Monitor, June 2015 - Transforming Urban Transport," Malaysia Economic Monitor 2015, World Bank Group, 2015. [Online]. Available: http://www.worldbank.org/en/country/malaysia/publication/malaysia-economic-monitor-june-2015. [Accessed: 04-Mar-2019].

[8] M. Ali, S. Manogaran, K. M. Yusof, and M. R. M. Suhaili, "Analysing vehicular congestion scenario in Kuala Lumpur using open traffic," Indones. J. Electr. Eng. Comput. Sci., vol. 10, no. 3, pp. 875–882, 2018.

[9] J. Ali, T. Ali, Y. Alsaawy, A. S. Khalid, and S. Musa, "Blockchain-based Smart-IoT Trust Zone Measurement Architecture," in Proceedings of the International Conference on Omni-Layer Intelligent Systems - COINS '19, 2019, pp. 152–157.

[10] Nielsen Insights, "Cash Or Cashless? Malaysia's Shifting Payment Landscape," Nielsen Malaysia, 2019. [Online]. Available: https://www.nielsen.com/my/en/insights/reports/2019/cash-or-cashless-malaysias-shifting-payment-landscape.html. [Accessed: 07-Jan-2019].

[11] N. M. Noor, S. M. Sam, N. Firdaus, M. Azmi, R. Che, and M. Yusoff, "RFID-based Electronic Fare Toll Collection System for Multi-Lane Free Flow – A Case Study towards Malaysia Toll System Improvement," J. Telecommun. Electron. Comput. Eng., vol. 8, no. 4, pp. 71–76, 2016.

[12] S. P, V. Valsan, and P. C, "IoT Based RFID Gate Automation System," Int. J. Eng. Trends Technol., vol. 36, no. 9, pp. 471–473, 2016.

[13] S. Nandhini and P. Premkumar, "Automatic Toll Gate System Using Advanced RFID and GSM Technology," vol. 3, no. 5, pp. 13002–13007, 2014.

[14] J. Ali, A. S. Khalid, E. Yafi, S. Musa, and W. Ahmed, "Towards a Secure Behavior Modeling for IoT Networks Using Blockchain," in CEUR Workshop Proceedings, 2019, pp. 244–258.

[15] C.-L. Hsu and J. C.-C. Lin, "Exploring Factors Affecting the Adoption of Internet of Things Services," J. Comput. Inf. Syst., vol. 58, no. 1, pp. 49–57, 2018.

[16] L. Gao and X. Bai, "A unified perspective on the factors influencing consumer acceptance of internet of things technology," Asia Pacific J. Mark. Logist., vol. 26, no. 2, pp. 211–231, 2014.

[17] K.-L. A. Yau, S. L. Lau, H. N. Chua, M. H. Ling, V. Iranmanesh, and S. C. C. Kwan, "Greater Kuala Lumpur as a smart city: A case study on technology opportunities," 2016 8th Int. Conf. Knowl. Smart Technol., no. February, pp. 96–101, 2016.

[18] S. F. Verkijika and L. De Wet, "E-government adoption in sub-Saharan Africa," Electron. Commer. Res. Appl., vol. 30, no. May, pp. 83–93, 2018.

[19] J.-S. C. Lin and P.-L. Hsieh, "Assessing the Self-service Technology Encounters: Development and Validation of SSTQUAL Scale," J. Retail., vol. 87, no. 2, pp. 194–206, Jun. 2011.

[20] S. M. Hizam and W. Ahmed, "A conceptual paper on SERVQUAL-framework for assessing quality of Internet of Things (IoT) services," Int. J. Financ. Res., vol. 10, no. 5, pp. 387–397, 2019.

[21] Y. Feng and Q. Xie, "Privacy Concerns, Perceived Intrusiveness, and Privacy Controls: An Analysis of Virtual Try-on Apps," J. Interact. Advert., vol. 2019, pp. 1–41, 2018.

[22] S. K. Sharma, S. K. Mangla, S. Luthra, and Z. Al-Salti, "Mobile wallet inhibitors: Developing a comprehensive theory using an integrated model," J. Retail. Consum. Serv., vol. 45, no. July, pp. 52–63, Nov. 2018.

[23] Krishna Kumar, Sivashanmugam C, and A. Venkataraman, "Intention To Use Mobile Wallet : Extension of Tam," Int. J. Curr. Eng. Sci. Res., vol. 4, no. 12, pp. 5–12, 2017.

[24] Z. Mani and I. Chouk, "Drivers of consumers' resistance to smart products," J. Mark. Manag., vol. 33, no. 1–2, pp. 76–97, 2017.

[25] T. Natarajan, S. A. Balasubramanian, and D. L. Kasilingam, "Understanding the intention to use mobile shopping applications and its in fl uence on price sensitivity," J. Retail. Consum. Serv., vol. 37, no. January, pp. 8–22, 2017.

[26] R. Agarwal and J. Prasad, "A Conceptual and Operational Definition of Personal Innovativeness in the Domain of Information Technology," Inf. Syst. Res., vol. 9, no. 2, pp. 204–215, 1998.

[27] M. Iranmanesh, S. Zailani, S. Moeinzadeh, and D. Nikbin, "Effect of green innovation on job satisfaction of electronic and electrical manufacturers' employees through job intensity: personal innovativeness as moderator," Rev. Manag. Sci., vol. 11, no. 2, pp. 299–313, 2017.

[28] X. Zhou, Q. Song, Y. Li, H. Tan, and H. Zhou, "Examining the influence of online retailers' micro-blogs on consumers' purchase intention," Internet Res., vol. 27, no. 4, pp. 819–838., 2017.

[29] M. Barth, P. Jugert, and I. Fritsche, "Still underdetected - Social norms and collective efficacy predict the acceptance of electric vehicles in Germany," Transp. Res. Part F Traffic Psychol. Behav., vol. 37, pp. 64–77, 2016.

[30] N. Huete-Alcocer, "A Literature Review of Word of Mouth and Electronic Word of Mouth: Implications for Consumer Behavior," Front. Psychol., vol. 8, no. July, pp. 1–4, 2017.

[31] F. D. Davis, R. P. Bagozzi, and P. R. Warshaw, "User Acceptance of Computer Technology: A Comparison of Two Theoretical Models," Manage. Sci., vol. 35, no. 8, pp. 982–1003, 1989.

[32] I. Ajzen, "The theory of planned behavior," Orgnizational Behav. Hum. Decis. Process., vol. 50, pp. 179–211, 1991.

[33] S. Pfoser, O. Schauer, and Y. Costa, "Acceptance of LNG as an alternative fuel: Determinants and policy implications," Energy Policy, vol. 120, no. April, pp. 259–267, 2018.

[34] B. Feng, Q. Ye, and B. J. Collins, "A dynamic model of electric vehicle adoption : The role of social commerce in new transportation," Inf. Manag., 2018.

[35] P. Taylor, E. Park, H. Kim, and J. Y. Ohm, "Understanding driver adoption of car navigation systems using the extended technology acceptance model," Behav. Inf. Technol., pp. 37–41, 2014.

[36] C.-K. Farn, Y.-W. Fan, and C.-D. Chen, "Predicting Electronic Toll Collection Service Adoption: An Integration of the Technology Acceptance Model and the Theory of Planned Behavior," Technol. E-government, 2007.

[37] H. Irawan, R. Hendayani, and D. Widyani, "Adoption of Electronic Toll Application Analysis," Int. J. Econ. Manag., vol. 10, no. S1, pp. 211–222, 2016.

[38] F. D. Davis, "Perceived Usefulness, Perceived Ease Of Use, And User Acceptance of Information Technology," MIS Quarterly; Sep, vol. 13, no. 3, pp. 319–240, 1989.

[39] A. A. Kervick, M. J. Hogan, D. O'Hora, and K. M. Sarma, "Testing a structural model of young driver willingness to uptake Smartphone Driver Support Systems," Accid. Anal. Prev., vol. 83, pp. 171–181, 2015.

[40] R. C. Jou, Y. C. Chiou, and J. C. Ke, "Impacts of impression changes on freeway driver intention to adopt electronic toll collection service," Transp. Res. Part C Emerg. Technol., vol. 19, no. 6, pp. 945–956, 2011.

[41] J. Globisch, E. Dütschke, and J. Schleich, "Acceptance of electric passenger cars in commercial fleets," Transp. Res. Part A Policy Pract., vol. 116, no. June, pp. 122–129, 2018.

[42] R. C. Jou, Y. C. Chiou, and J. C. Ke, "Impacts of impression changes on freeway driver intention to adopt electronic toll collection service," Transp. Res. Part C Emerg. Technol., 2011.

[43] M. J. Alsamydai, R. O. Yousif, and M. H. Al-Khasawneh, "The Factors Influencing Consumers ' Satisfaction and Continuity to Deal With E-Banking Services in Jordan," Glob. J. Manag. Businesss Res., vol. 12, no. 14, 2012.

[44] B. Alonso, R. Barreda, L. Olio, and A. Ibeas, "Modelling user perception of taxi service quality," Transp. Policy, vol. 63, pp. 157–164, 2018.

[45] A. Parasuraman, Z. Valarie A., and B. Leonard L., "Servqual : A Multiple-Item Scale For Measuring Consumer Perc," J. Retail., vol. 64, p. 12, 1988.

[46] A. Parasuraman, V. A. Zeithaml, and A. Malhotra, "E-S-QUAL a multiple-item scale for assessing electronic service quality," J. Serv. Res., vol. 7, no. 3, pp. 213–233, 2005.

[47] M. S. Janita and F. J. Miranda, "Quality in e-Government services: A proposal of dimensions from the perspective of public sector employees," Telemat. Informatics, vol. 35, no. 2, pp. 457–469, 2018.

[48] S. Hsu, F. Qing, C. Wang, and H. Hsieh, "Evaluation of service quality in facebook-based group-buying," Electron. Commer. Res. Appl., vol. 28, pp. 30–36, Mar. 2018.

[49] M. J. Alsamydai, "Adaptation of the Technology Acceptance Model (TAM) to the Use of Mobile Banking Services," Int. Rev. Manag. Bus. Res., no. 2014, pp. 2016–2028, 2016.

[50] E. Considine and K. Cormican, "Self-service Technology Adoption: An Analysis of Customer to Technology Interactions," in Procedia Computer Science, 2016, vol. 100, pp. 103–109.

[51] S. Boon-itt, "Managing self-service technology service quality to enhance e-satisfaction," Int. J. Res. Dev., vol. 7, no. 1, pp. 63–83, 2016.

[52] H. Boeck, J. Roy, F. Durif, and M. Grégoire, "The effect of perceived intrusion on consumers ' attitude towards using an RFID-based marketing program," vol. 5, pp. 841–848, 2011.

[53] Z. Mani and I. Chouk, "Drivers of consumers' resistance to smart products," J. Mark. Manag., vol. 33, no. 1–2, pp. 76–97, 2016.

[54] V. M. Wottrich, E. A. van Reijmersdal, and E. G. Smit, "The privacy trade-off for mobile app downloads: The roles of app value, intrusiveness, and privacy concerns," Decis. Support Syst., vol. 106, pp. 44–52, 2018.

[55] S. Hérault and B. Belvaux, "Privacy paradox et adoption de technologies intrusives Le cas de la géolocalisation mobile," Décisions Mark., no. 74, pp. 67–82, 2014.

[56] J. Lu, J. E. Yao, and C. S. Yu, "Personal innovativeness, social influences and adoption of wireless Internet services via mobile technology," J. Strateg. Inf. Syst., vol. 14, no. 3, pp. 245–268, 2005.

[57] R. Rothwell, Towards the Fifth‐generation Innovation Process, vol. 11, no. 1. 1994.

[58] K. ; Koivisto et al., "Extending the Technology Acceptance Model with Personal Innovativeness and Technology Readiness : A Comparison of Three Models," vol. 22, pp. 113–128, 2016.

[59] A. Turan, A. Ö. Tunç, and C. Zehir, "A Theoretical Model Proposal: Personal Innovativeness and User Involvement as Antecedents of Unified Theory of Acceptance and Use of Technology," Procedia - Soc. Behav. Sci., vol. 210, pp. 43–51, 2015.

[60] Z. Lin and R. Filieri, "Airline passengers' continuance intention towards online check-in services: The role of personal innovativeness and subjective knowledge," Transp. Res. Part E Logist. Transp. Rev., vol. 81, pp. 158–168, 2015.

[61] Chu Shu-chuan and Kim Yoojung, "Determinants of consumer engagement in electronic word-of-mouth (eWOM) in social networking sites," Int. J. Advert., vol. 30, no. 1, pp. 47–75, 2011.

[62] N. B. Ellison and D. M. Boyd, Sociality Through Social Network Sites, vol. 1. Oxford University Press, 2013.

[63] M. Soderlund and S. Rosengren, "Receiving word-of-mouth from the service customer: An emotion-based effectiveness assessment'," J. Retail. Consum. Serv., vol. 14, no. 2, pp. 123–136, 2007.

[64] U. Gretzel and K. H. Yoo, "'Use and Impact of Online Travel Reviews,'" in Information and Communication Technologies in Tourism, 2008, pp. 35–46.

[65] A. Chong and E. Ngai, "What Influences Travellers' Adoption of a Location-based Social Media Service for Their Travel Planning?," in PACIS 2013 Proceedings, 2013.

[66] M. Ajzen, I. Fishbein, "Understanding Attitudes and Predicting Social Behavior. Prentice-Hall Inc., Englewood Cliffs, NJ.," Prentice-Hall Inc., Englewood Cliffs, NJ., 1980.

[67] M. N. K. Saunders, P. Lewis, and A. Thornhill, Research methods for business students, 7th ed. 2015.

[68] G. C. Moore and I. Benbasat, "Development of an Instrument to Measure the Perceptions of Adopting an Information Technology Innovation Stable URL : http://www.jstor.org/stable/23010883 Linked references are available on JSTOR for this article : of an Instrument to Measure the Percepti," Inf. Syst. Res., vol. 2, no. 3, pp. 192–222, 1991.

[69] R. E. Goldsmith and C. F. Hofacker, "Measuring consumer innovativeness," J. Acad. Mark. Sci., vol. 19, no. 3, pp. 209–221, Jun. 1991.

[70] S. Bambauer-Sachse and S. Mangold, "Brand equity dilution through negative online word-of-mouth communication," J. Retail. Consum. Serv., vol. 18, no. 1, pp. 38–45, 2011.

[71] I. Ajzen, "The Theory of Planned Behavior," 1991.

[72] World Population Review, "Kuala Lumpur Population 2018 (Demographics, Maps, Graphs)," World Population Review. 2018.

[73] M. N. K. Saunders, P. Lewis, and A. Thornhill, Research methods for business students. Pearson, 2012.

[74] J. F. Hair, M. Sarstedt, C. M. Ringle, and J. A. Mena, "An assessment of the use of partial least squares structural equation modeling in marketing research," J. Acad. Mark. Sci., vol. 40, no. 3, pp. 414–433, 2012.

[75] Y. Owari and N. Miyatake, "Prediction of chronic lower back pain using the hierarchical neural network: Comparison with logistic regression—A pilot study," Med., vol. 55, no. 6, 2019.

[76] R. N. D'souza, P.-Y. Huang, and F.-C. Yeh, "Small Data Challenge: Structural Analysis and Optimization of Convolutional Neural Networks with a Small Sample Size," bioRxiv, p. 402610, 2018.

[77] J. F. Hair., W. C. Black, B. J. Babin, and R. E. Anderson, Multivariate Data Analysis William C . Black Seventh Edition. 2014.